# Sustainable restoration of intermittent streams: Integrating ecological design and urban resilience

Parinaz Baradaran Anaraki*, Shiva Manshour

West Virginia University, Morgantown, West Virginia, 26506, USA
University of Nevada, Las Vegas, Las Vegas, Nevada, 89154, USA
PB00030@MIX.WVU.EDU, shiva.manshour@unlv.edu

## Abstract

The sustainable restoration of intermittent streams has become a critical priority in contemporary urban planning, particularly as cities confront the dual challenges of ecological degradation and climate change. In Tehran, decades of rapid urbanization and poor management practices have confined natural streams into rigid concrete channels, eroding their ecological value and disconnecting them from community life. This paper introduces an **ecological and sustainability-oriented framework** for the restoration of the Darband and Darabad river valleys, highlighting their potential to function as ecological corridors that support **biodiversity, thermal regulation, cultural identity, and urban resilience**.

The study employs a systematic methodology that integrates ecological engineering, landscape design, hydrological modeling, and participatory planning. Findings suggest that restoring these river valleys through sustainable strategies, such as the creation of active green networks, multifunctional public spaces, and resilient hydrological systems, can transform them from degraded drainage corridors into **life-giving urban landscapes**. Moreover, the research emphasizes the necessity of linking restoration with sustainability goals to ensure long-term ecological balance, social well-being, and climate adaptation.

This case study demonstrates that sustainable river restoration, when aligned with ecological design and community engagement, has the potential to reposition intermittent streams as essential infrastructures for **sustainable urban development and resilience**.

## Introduction

The rivers of Tehran, particularly **Darband** and **DarAbad**, present unique circumstances. Historically, these rivers had only a negligible impact on their surroundings, largely because the city's primary water supply relied on a traditional network of qanats [1]. Long before their canalization, they were perceived as barriers to urban expansion rather than as vital ecological and cultural lifelines. Today, however, in the context of climate change and escalating environmental risks, these rivers must be recognized as critical natural assets with the potential to contribute to sustainability and resilience.

Despite this importance, unbalanced urban development and the failure to embrace the ecological value of natural resources have disrupted the delicate balance between built and natural systems.

The gradual retreat of urban rivers, increasingly surrendered to rigid concrete structures, has become both a visible scar and an ecological loss. In recent decades, uncoordinated practices and the destructive approach of eliminating rivers from development processes have deprived both citizens and urban managers of their multiple benefits. This reductive trend has gone so far as to convert river corridors into unauthorized sewage networks. Yet, urban rivers are widely acknowledged as having the potential to **restore balance between the built environment and natural systems**, reinforcing ecological integrity and community well-being [2].

One of the most prominent forms of water's presence in Tehran's urban structure is its system of **river valleys**, still discernible despite extensive construction and road networks. Flowing from the Alborz Mountains to the south, these valleys historically served as connecting corridors between the city and nature. Their significance today extends beyond ecological value, encompassing historical, cultural, and environmental dimensions. While each of Tehran's main river valleys has developed distinct spatial, morphological, and functional characteristics under current urban pressures, they remain interconnected as part of a broader river network. With appropriate planning and sustainable strategies, this system can guarantee greater spatial and ecological integration of the city [3].

For this reason, Tehran's **new master plan** emphasizes coordinated interventions and an integrated approach to its river valleys, designating five of the seven primary valleys (Kan, Farahzad, Darake, Darband, and DarAbad) as central pillars of sustainable urban development.

## Methodology and Findings

Although many scholars have examined the influence of river valleys on urban life, there remains a lack of formal knowledge regarding how large-scale changes have specifically shaped them in Iran. The beginning of rehabilitation efforts in the 1990s, following the Iran-Iraq war, coincided with the urgent need for reconstruction and mass development to compensate for war-related damages. This rapid transformation deeply affected Iran's culture and industry, while also reflecting the broader geopolitical context and the internationalization of architecture influenced by modernism.

The purpose of this study is to assess whether such analysis can reveal potential competition and synergy between human-made architecture and natural systems. To address these questions, a **systematic literature review** was conducted, analyzing research published between 1959 and 2021. This 60-year overview highlights current research trends, identifies gaps in the existing literature, and underscores the overlooked potential for rehabilitating river valleys and their surrounding areas.

A combination of diverse ideologies, social pressures, and environmental challenges, along with the absence of academic guidelines for river rehabilitation, has contributed to the marginalization of these natural systems in architectural and planning discourses. One major limitation of this study lies in the scarcity or distortion of resources due to political and social constraints, making the identification and extraction of credible sources particularly demanding.

To regenerate Tehran's **vitality network**, the first step is to understand its nodes and points of connection. A general typology was developed to categorize nodes related to the river valleys, based on the following principles:
- Analysis of land ownership patterns in urban zones

- Market tendencies influencing local interventions (approved development plans, ongoing projects, investor interest)
- Physical and spatial characteristics of each location
- Functional activities and their spheres of influence
- Barriers, as well as economic and developmental drivers
- Policies and regulations shaping the direction of urban development

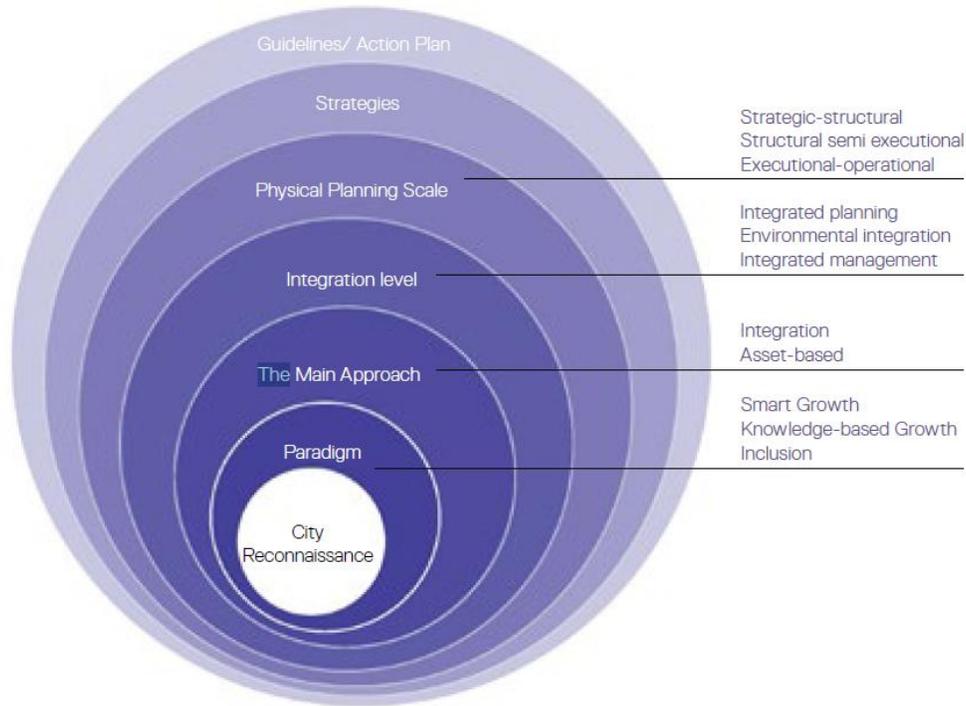

**Figure. 1**: Strategic approach

# Literature Review

**Site Analysis**
Human industrialization has led to a significant decline in river quality worldwide. Drivers such as population growth, rapid urbanization, and climate change are expected to place increasing pressure on river ecosystems in the coming decades. When river systems degrade to such an extent that they can no longer provide essential ecological services, the concept of **river restoration** becomes indispensable [4].

Similarly, the integration of ecological and cultural assets into urban design echoes findings from [5], which demonstrated that traditional passive cooling strategies in hot-arid regions, such as courtyards, wind catchers, shading systems, thermal mass, and vegetation form climate-responsive systems that harmonize human comfort with natural processes. This analogy reinforces the potential of Tehran's river valleys to function as ecological corridors that provide thermal regulation, cultural identity, and sustainable urban resilience.

Tehran contains several river valleys that run from north to south, in addition to a number of canals and seasonal streams that absorb excess rainfall. Within the city's environmental planning framework, some river valleys have undergone partial regeneration, while others have been designated for urban greens and gardens. These strategies have been recognized as important responses to Tehran's pressing urban challenges, including severe pollution, social insecurity, and the shortage of accessible cultural spaces.

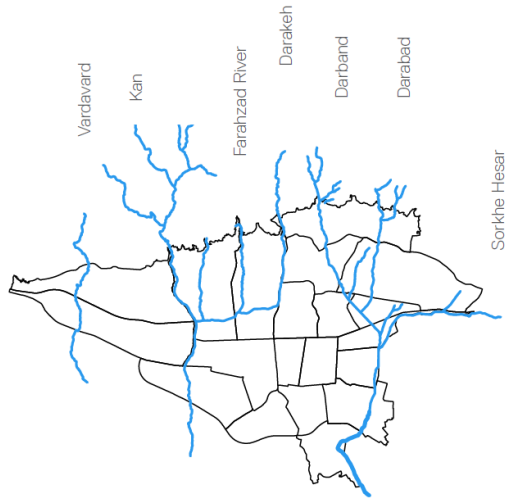
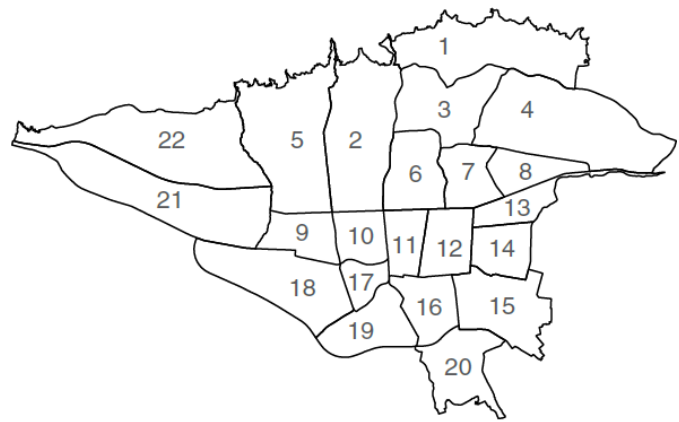

**Figure. 2**: Tehran's Rivers                    **Figure. 3**: Tehran's regions

**Darband River-Valley Analysis**

The **Darband River Valley**, stretching 33 km with an elevation difference of about 1 km between its urban and peri-urban zones, extends from the northern Alborz Mountains to the southernmost edges of Tehran. It passes through six urban districts with diverse socio-economic and environmental conditions, continues across the plains, and ultimately reaches the desert areas of southeast Tehran. In the southern sector, beyond the river, lies a semi-mountainous area known as *Bibi Shahr Banoo*.

To date, the most significant intervention along this 33 km corridor has been the creation of a **mountainous recreational area** in the northern section. In contrast, much of the remaining valley has not been integrated into urban planning or design projects. Within the city, the river flows largely through covered and open concrete canals; further south, in the peri-urban areas, it continues past several major parks before stretching into the dry lands of central Iran [6].

The rivers and streams traversing these mountain valleys remain among Tehran's most valuable **landscape assets**, offering strong ecological potential to create vital green patches within the dense urban fabric. Beyond their ecological role, they also provide open spaces, define key visual corridors, and facilitate natural air circulation. For these reasons, the city's **comprehensive landscape plan** identifies these natural corridors as major components for restoring Tehran's ecological vitality. By enabling the creation of continuous and expansive natural patches and ensuring spatial integration between green and built environments, the river valleys are regarded as central pillars of Tehran's **sustainability development plan** [6].

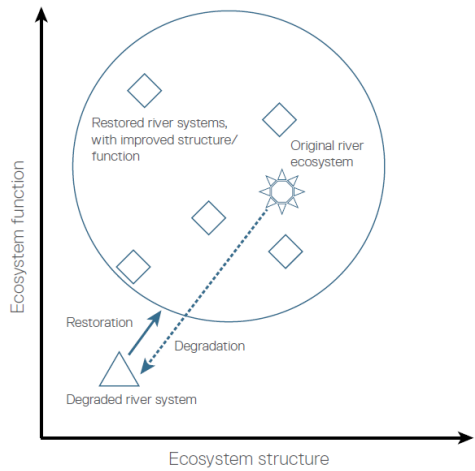

**Figure. 4**: Tehran's ecosystem

These river valleys follow different trajectories across Tehran. The **Kan River Valley** extends further south until it reaches the central plains. In contrast, the **Darke** and **Farahzad** river valleys terminate in the western parts of the city. The **DarAbad** and **Sokhe Hesar** valleys are redirected into the Abouzar canal, while the **Darband River Valley**, after passing through the Maghsoud Beyk and Bakhtar watercourses, also eventually connects to this canal. The spatial distribution and connections of these river valleys within Tehran are illustrated in the maps provided (Figure 4).

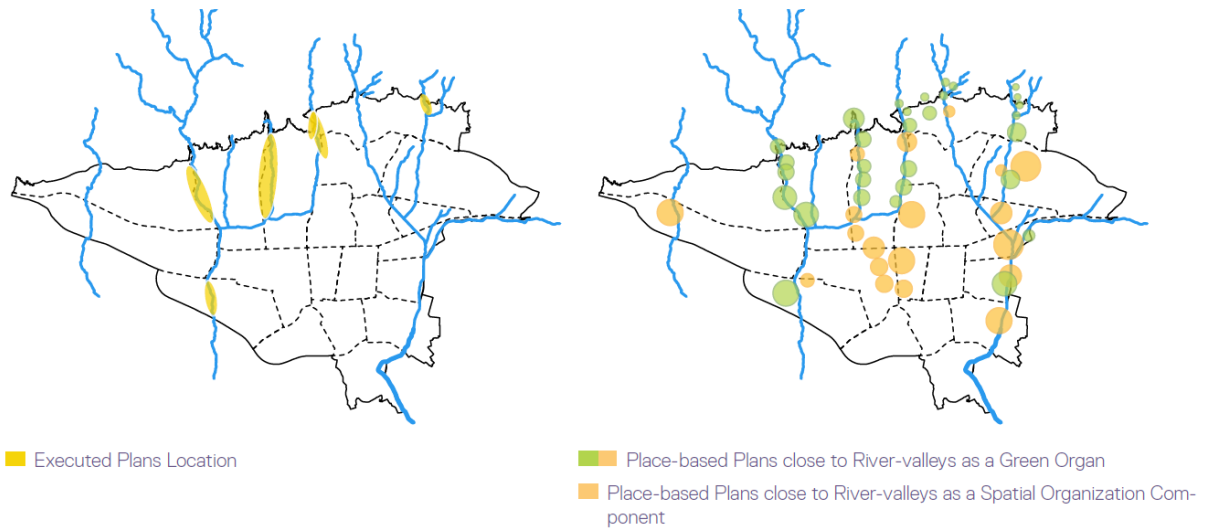

**Figure. 5**: The position of Tehran's rivers

The study focused on the entire stretches of the **Darband**, **Golab Dare**, and **Dar Abad** rivers, along with their surrounding areas, covering approximately 40 km and including the eastern flood canal (*Masil Bakhtar*). These river valleys are predominantly surrounded by green spaces, which provide significant ecological and landscape potential. The investigation area extended 100 to 150 meters on both sides of each river axis, integrating both principles of river engineering and perspectives from landscape design and environmental planning.

Within the city limits, the adjacent urban fabric along these rivers displays diverse morphological patterns. This area, encompassing approximately 2,300 hectares, represents the subject of detailed urban design studies aimed at achieving sustainable integration between the built environment and natural river systems.

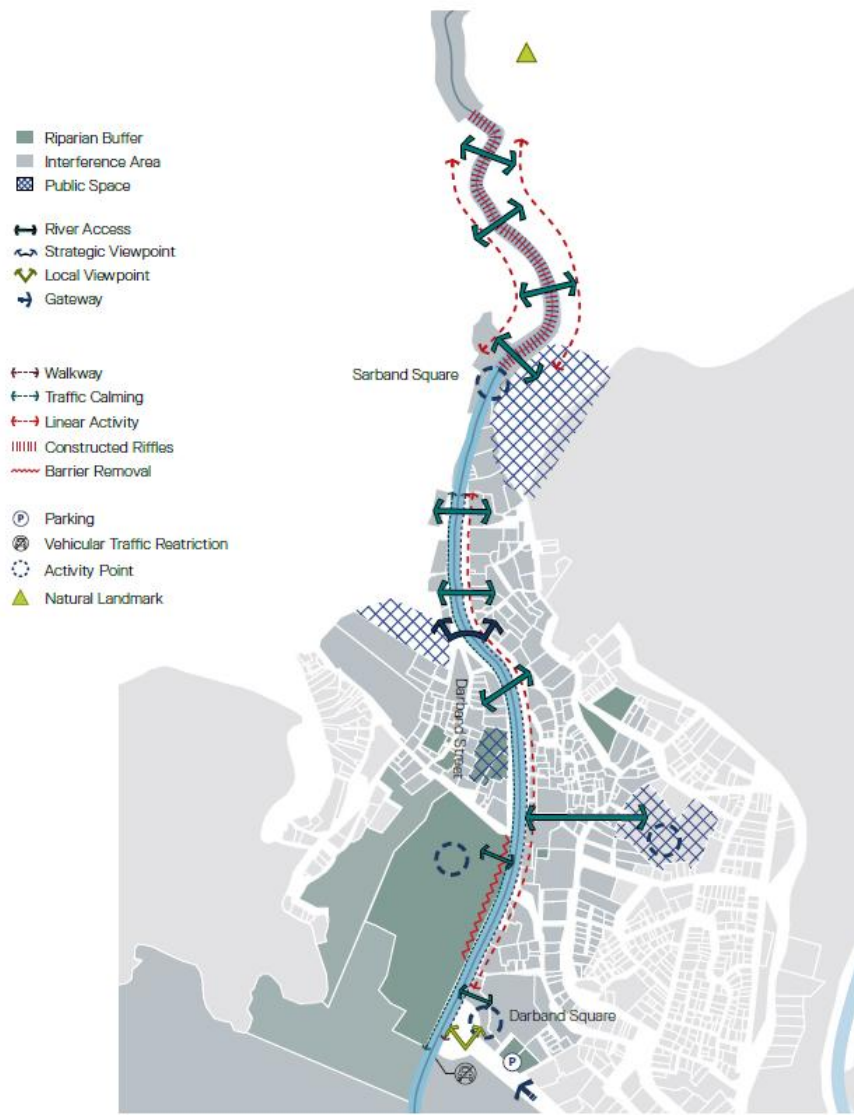

**Figure. 6**: Darband and Darabad site plan

**Problem Statement**

The rivers and natural streams of Tehran have been subjected to continuous damage, both in their ecological origins and in their urban identity, as a result of human activity and weak urban management. Historically, these waterways served a vital purpose: channeling surface waters from snowmelt, rainfall, and mountain springs in the northern and eastern highlands into the city, where they provided resources for irrigation, drinking, and ecological balance. Today, however, they have largely been transformed into **drainage systems for polluted urban waters**, functioning as conduits for sewage, garbage, and other imposed elements of the city.

Due to their foul odor, pollution, and the spread of insects and parasites along their banks, many of these rivers have been confined to covered canals in central and southern Tehran, or redirected into tunnels and flood embankments to isolate them from the urban environment as much as possible. This treatment has alienated the rivers from the life of the city.

Yet, water is universally recognized as a **fundamental element of vitality, energy, and mobility in cities**. The prevailing perception of rivers as obstacles to development must change. What remains of these natural resources should be redefined as **life-giving ecological assets** rather than drainage channels. The transformation of their identity must begin with shifting away from their current role as polluted drains and toward their **restoration based on principles of urban planning and ecological sustainability**, ensuring that their crucial role in nature and urban life can be preserved [7].

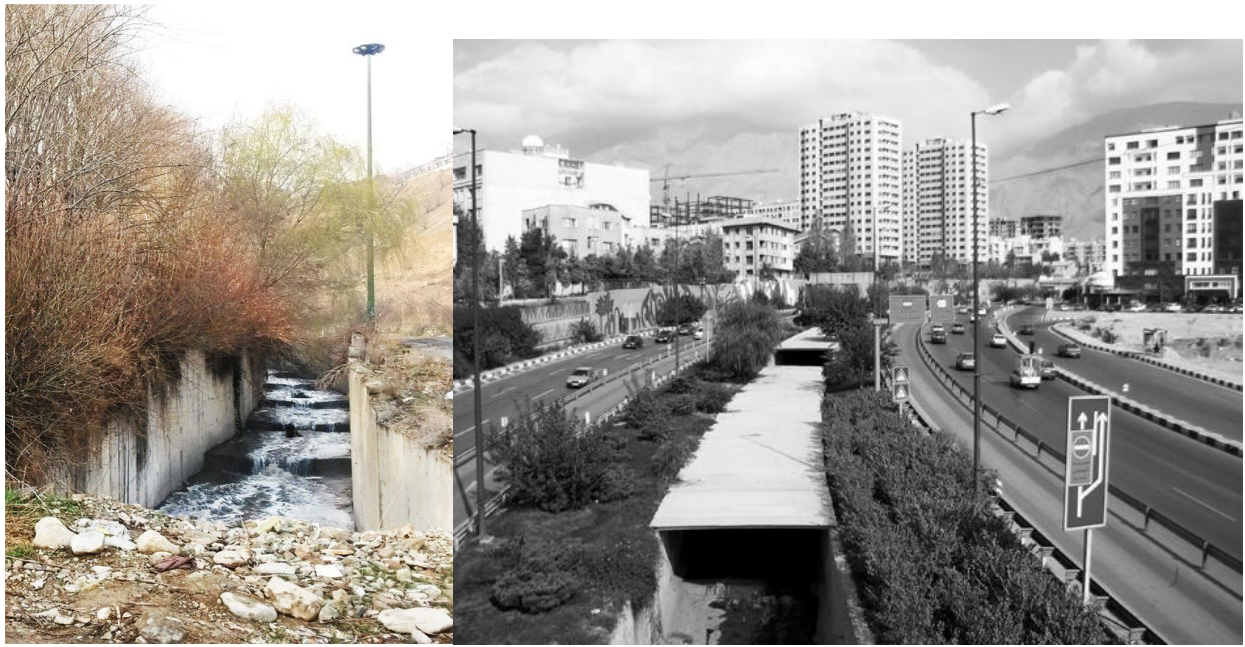

**Figure. 7**: The current situation of Tehran's rivers

**Restoration Context and Challenges**

River restoration is influenced by a wide range of factors, including ecological conditions, social and cultural dynamics, economic constraints, the protection of infrastructure and assets from water-related risks, and, increasingly, the global imperative of **water preservation**. As an integral part of water resources management systems, river restoration plays a vital role in balancing human demand for freshwater ecosystem services with the anthropogenic pressures that degrade river ecosystems. Achieving this balance requires a clear understanding of the relationship between the river's **ecological functions** and the diverse demands placed upon them.

Over time, restoration practices have evolved from one-dimensional responses targeting single issues (such as water quality) to more **comprehensive approaches** that address multiple challenges simultaneously. These approaches now encompass both **active restoration**, which involves physical interventions to modify rivers and surrounding landscapes, and **passive restoration**, which relies on policy frameworks and management measures to conserve and protect existing resources [8].

Despite these advancements, river policy managers continue to face significant challenges in restoring river ecosystems [4]:
1. Returning rivers to their original, pre-disturbance state is rarely feasible.
2. Restoration efforts must reconcile the multiple and often competing roles of rivers.
3. The inherent complexity and scale of river systems make effective restoration planning and implementation difficult.
4. Growing uncertainty about future environmental and climatic conditions complicates long-term restoration efforts.

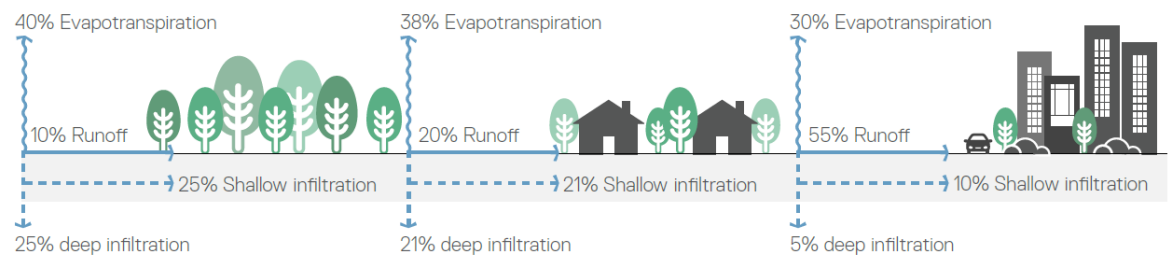

**Figure. 8**: How Urbanization threatens Natural water exchange processes in Tehran

**Strategies**

Tehran's **vitality network** is envisioned as a comprehensive urban framework that integrates the built environment, economy, and public spaces to provide a new roadmap for the city's sustainable future. This network aims to restore the liveliness of Tehran through consistent and integrated management, with a particular emphasis on strengthening **human–nature interactions**. Its components, supported by spatial flexibility, converge existing facilities to create new opportunities for interaction between citizens and the city's evolving ecological systems. In this sense, the vitality network establishes a renewed identity for Tehran based on a contemporary definition of urbanization [9].

In general, the vitality network is composed of multiple layers, structured around two primary elements: **connection points** and **nodes**.
- **Connection points** fall into two categories: river valleys and urban networks. These either already contain natural and green structures or hold the potential to incorporate additional flexible green spaces.
- **Nodes** can be categorized into several groups [10]:

- Large-scale green spaces and open fields that can serve as key nodes within the vitality network (e.g., Pardisan Park, Ghale Morghi, Abas Abad Heights).
- Urban activity zones, particularly those dedicated to tourism, recreation, and leisure, which, when combined with flexible activities and green spaces, become critical pillars of the network.
- Large-scale facilities or complexes whose spatial structure can be redefined and merged with the vitality framework to support the regeneration of Tehran's natural environment. Examples include Imamzade shrines, prisons, hospitals, airports, cemeteries, hotels, large residential complexes such as Atisaz, and sports complexes like Enghelab and Ararat.
- Suburban zones that possess significant natural or cultural identity.
- Public spaces such as Tajrish Bazaar, which function as diverse urban nodes within the network, integrating multiple functions and cultural dimensions [7].

The characteristics of these nodes are vital for recreating Tehran's vitality network, as each reflects a distinct aspect of the city's identity. Unlike the city's current rigid and fragmented urban structure, the vitality network offers **flexibility, adaptability, and integration**.

Tehran's **river rehabilitation strategic plan** is aligned with this framework, aiming to manage these layers in an integrated manner and harmonize the development of all components, including nodes and connection points, into a cohesive system. Ultimately, this approach is expected to regenerate Tehran's vitality network and support its transition toward sustainable urban development [11].

The **values and strategies** of the restoration plan for the Darband and Dar Abad rivers are summarized in the following table.

| VALUES | STRATEGIES |
| --- | --- |
| SECURITY OF THE BUILT ENVIRONMENT AND THE RIVER | A. River Security on account of Hydraulic and Flood Parameters<br>B. River Security on account of Morphology and Erosion Parameters<br>C. River Security from the Perspective of Geology Parameters<br>D. Studying the Bed Limits and Borders of the River and the Assaults and Complaints, and Differences in this Matter |
| THE RIVER'S ELEMENTS | A. Water Quality<br>B. The Pollution and Purifier Sources of the River |
| INTEGRATION OF THE RIVER AND THE BUILT ENVIRONMENT | A. Identifying Development Opportunities<br>B. Merging the built environment with the river |
| IDENTITY (RIVERS, CITY IMAGE AND SENSE OF PLACE) | A. River's Qualitative Character improvement<br>B. Public Access improvement<br>C. Public Spaces |
| CONSERVATION AND INCREASE OF THE THE CONTEXT FOR CREATING URBANIZED ECOSYSTEMS | A. Improving Green spaces<br>B. Improving the Structures of Physical Habitats |

|  | C. Identifying the River's potential |
| --- | --- |
| THE WATER FLOW OF THE RIVER | A. Water resources identification and their Accessibility and Future<br>B. Analyzing water consumption |
| LOCATION OF AQUIFERS AND GROUNDWATER | A. Improving the Condition of Aquifers<br>B. Analyzing the Condition of Springs and Qanats<br>c. Analyzing the Evacuation and Removal Points of Water |
| AWARENESS OF DECISION-MAKING GROUPS AND ORGANIZATIONS | A. Creating data for Supporting the System's Projects and Subprojects<br>B. Running workshops to Transfer Experience |

**Table 1**: Values and strategies [12].

**Darband and Darabad Restoration Planning Process**

For the past three decades, the predominant treatment of urban river streams in Tehran has been limited to channelization, a practice that has altered hydrological processes and widened the gap between rivers and surrounding neighborhoods. This has resulted in both ecological degradation and a visual disconnection between the built environment and natural landscapes.

To address these issues, the Tehran municipality has adopted a new management perspective that envisions a sequence of thematic green spaces serving as meeting points for adjacent neighborhoods. Pilot projects have demonstrated that effective river restoration cannot rely on single-purpose interventions, but instead requires a holistic approach, supported by a range of technologies and planning tools aligned with integrated urban and ecological strategies.

The current framework for the Darband and Darabad catchments in northeast Tehran creates opportunities to restore natural systems and improve urban amenity, ecological health, and long-term sustainability. This framework is guided by four key planning principles:

- Reinforcing the city's unique relationship to the river through riverfront design that reflects ecological and cultural values.

- Understanding the river ecosystem and planning at a scale larger than the immediate riverfront to ensure landscape connectivity.

- Minimizing new floodplain development, thereby reducing risks and safeguarding natural hydrological functions.

- Expanding public access and connectivity, integrating recreational uses to strengthen human‑nature interactions.

The strategic plan is expected to provide Tehran with the capacity to assess and project the ecological and social impacts of future development within the catchment area. In doing so, it will help prevent interventions that could cause long-term ecological harm and instead foster a model of sustainable urban-river integration. unintended long-term impacts [12].

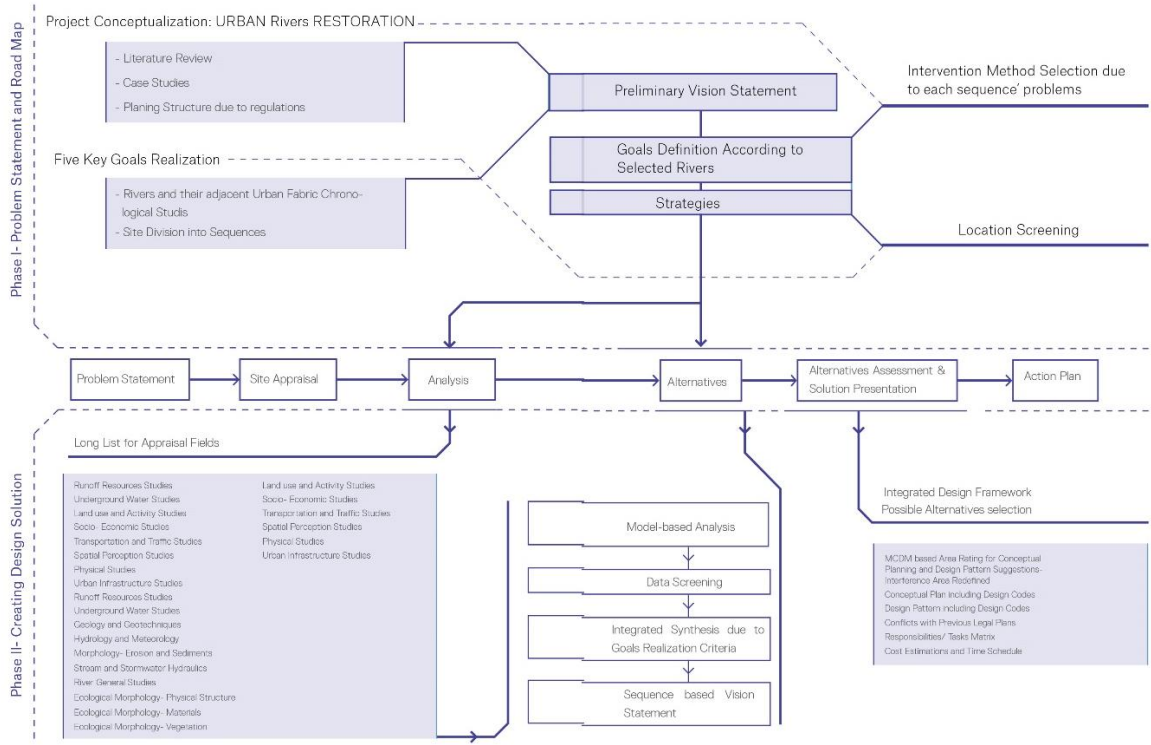

**Diagram 1**: Darband and Darabad restoration planning

**Challenges and opportunities**

To define a master plan for Darband and Darabad restoration, the potential for development in different structural, physical, and spatial factors was analyzed. Also, the economic, organizational, and natural resources were examined. Furthermore, at this level, field observation emphasized developing possibilities of the current situation. The challenges and opportunities can be categorized into five sections [12]: 1- Natural potentials 2- The economy 3- Function 4- The Physical environment 5- Transportation nodes and system

- *Natural potentials challenges:*

1- The presence of old trees around the intermittent zone as a green line provides a great potential for life and shading to reduce the water temperature
2- Parks and green local spaces as a part of the qualitative boundary of intermittent rivers
3- Lack of construction around the riverbed

- *Natural potentials Opportunities:*

1- Unbalanced dispersion of the green spaces in the area and their shortage in the north-eastern part
2- Using River borders as vehicular lines and construction up to the Riverhead
3- Poor ecosystem due to human interventions
4- Pollution of the rivers
5- Potentials for the floods due to the slopes

- *The economic challenges:*

1- High life expectancy level
2- Relative abundance of wealthy social classes
3- Higher literacy level than other corresponding areas in Tehran
4- The presence of various markets, shopping centers, bazaars, and activity centers
5- The presence of residential-garden areas with a low-density population in the district

- *The economic opportunities:*

1- Increasing the daytime activities
2- Increasing the density of the population
3- Increasing pedestrians in daytime
4- The change in economic-social classing because of migration

- *The Function challenges:*

1- The presence of residential neighborhoods around the river
2- Diversity in activities
4- The presence of active public spaces such as markets
5- Establishment of functional and activity nodes in important intersections
6- Establishment of active uses
7- Establishing residential areas to increase the public presence of the space
8- The opportunity for increasing commercial and recreational uses around the intermittent river
9- The opportunity for increasing green and recreational spaces

- *The Function opportunities:*

1- Lack of high-quality public spaces
2- The presence of unsuitable and incompatible activities (such as automobile services, educational and health centers) next to crowded spaces
3- Limited leisure and recreational spaces

- *The Physical environment challenges* [15].

1- Shallow enclosure of the river from Haghani to Hemmat highway
2- Diversity in the enclosure of public spaces
3- The opportunity for infill development in abandoned spaces or ruined buildings
4- The opportunity for the creation of architectural modules, articulating the building façades

5- The opportunity for a physical emphasis in corners and important intersections in terms of spatial-perceptual organization
6- The opportunity for increasing the functional diversity and physical interchangeability
7- Conservation and creation of sightseeing

- *The Physical environment opportunities:*

1- Lack of clear entrances
2- Lack of physical identification in nodes and important intersections
3- The presence of decayed urban facades
4- Low physical flexibility of buildings for the establishment of diverse uses
5- Visual disarray in the skyline and background, and the decayed urban facades
6- The abandoned spaces
7- Low permeability in the adjacent area of the river

- *Transportation nodes and system challenges:*

1- Adjacency of the river to railroad transportation hubs and the possibility of ease of access for pedestrians
2- Notable role of Shari'ati and Pasdaran axes as important arteries in the movement of individuals in the passage network system
3- Aligned direction of the rivers and passages
4- Connection with other commercial and residential areas through highways
5- The presence of different intersections with important artery axes and the creation of activity nodes and landmarks
6- The possibility of connecting the pedestrian paths through the surrounding areas of the rivers
7- The opportunity for controlling the vehicular movement based on the vehicle type
8- The potential for easy accessibility to hubs and public transportation stations

- *Transportation nodes and system opportunities:*

1- Prevailing dominance of the vehicle and decreasing presence of the pedestrian
2- Lack of adequate accessibility from rivers
3- Improper positioning of transportation stations such as taxis, minibusses, and generating visual, noise, and traffic pollutants
4- Disrupting the connectivity of pedestrian paths with highways
5- Illegibility of the intersections of intermittent streams with artery axes
6- Consigning the traffic nodes of the residential areas
6- The high slope from north to south makes it difficult for pedestrians and cyclists to pass

**Analyzing data**

Generally, the study includes five levels of process in the form of a hierarchical spatial structure of integrated restoration: Macro level (including Tehran city and the drainage basin of Darband intermittent river), site plan (including social, physical and spatial, hydrological, physiographical, and natural resources), intervention area (homogenous areas for urban planning, natural, and public participation), intervention nodes (spots with special opportunities for riparian restoration) and master plan [14].

1. *Macro level:*
- Analysis of the hydrological results of the upstream waters, including the drainage basins of Golabdareh and Darband and the river of Maqsudbeyk, the Drainage basin of Jamshidieh and Velenjak
- Analysis of the upstream plans of the urban planning and natural areas

2. *Site plan:*
- Categorizing the river, its size, and geometry
- Analysis of the morphological changes in the river
- Analysis of the river pollution based on the available information
- Defining the natural bed context and border of the river based on the border limit
- Library studies and field observations

3. *Intervention area:*
- A field study of the river features
- Conducting a public participation plan
- Establishing workshops for identifying homogeneous areas in terms of morphology, flow type, materials, land uses, physical-spatial, and emotional
- Integrated multi-disciplinary decision-making groups

4. *Intervention nodes:*
- Positioning of activity nodes

5. *Master plan:*
- Providing short-term, medium-term, and long-term concepts
- Evaluating the environmental results of the projects

**Technical Alternatives and Solutions**

Within this methodology, the rehabilitation of river valleys and the development of their surrounding areas were evaluated through a variety of technical alternatives. Multiple approaches, including model simulations, multi-objective optimizations, and diverse rehabilitation strategies, were analyzed, assessed, and ranked according to the required budgetary resources needed for implementation. This process provides decision-makers with the flexibility to establish land-management policies at varying levels of investment, ensuring adaptability to financial and institutional constraints [14].

In the subsequent stage, each decision was linked to the broader development trajectory of the river-valley areas. Different functional transformation methods were introduced, ultimately informing the design of the final master plan. To ensure comprehensive evaluation, the selection criteria were grouped into four overarching categories:

1. Spatial criteria – addressing land use, connectivity, and integration within the urban fabric.
2. Economic criteria – evaluating cost-effectiveness, investment priorities, and long-term value creation.

3. Social criteria – emphasizing community needs, accessibility, cultural identity, and human–nature interactions.
4. Environmental criteria – ensuring ecological sustainability, biodiversity protection, and resilience to climate change.

Through this structured approach, the methodology establishes a balanced framework that integrates ecological, social, and economic dimensions, guiding the sustainable restoration of river valleys and their urban contexts.

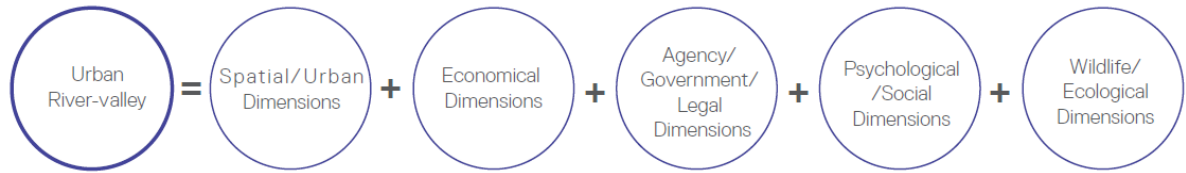

**Diagram.2**: The relationship between the components of the river [17].

During the decision-making stage, the plans and alternatives identified in the previous phase, based on spatial, economic, social, and environmental criteria, were systematically evaluated and ranked by managers and stakeholders. This participatory evaluation ensured that multiple perspectives were integrated into the process, balancing technical feasibility with community priorities.

Ultimately, the **land-management policies** for the river valleys were determined using a **multi-criteria decision-making (MCDM) model**, which allowed for transparent comparison of alternatives and the selection of strategies that best aligned with sustainability objectives.

The entire decision-making framework, including the ranking process and integration of stakeholder input, is illustrated in the following diagram [16].

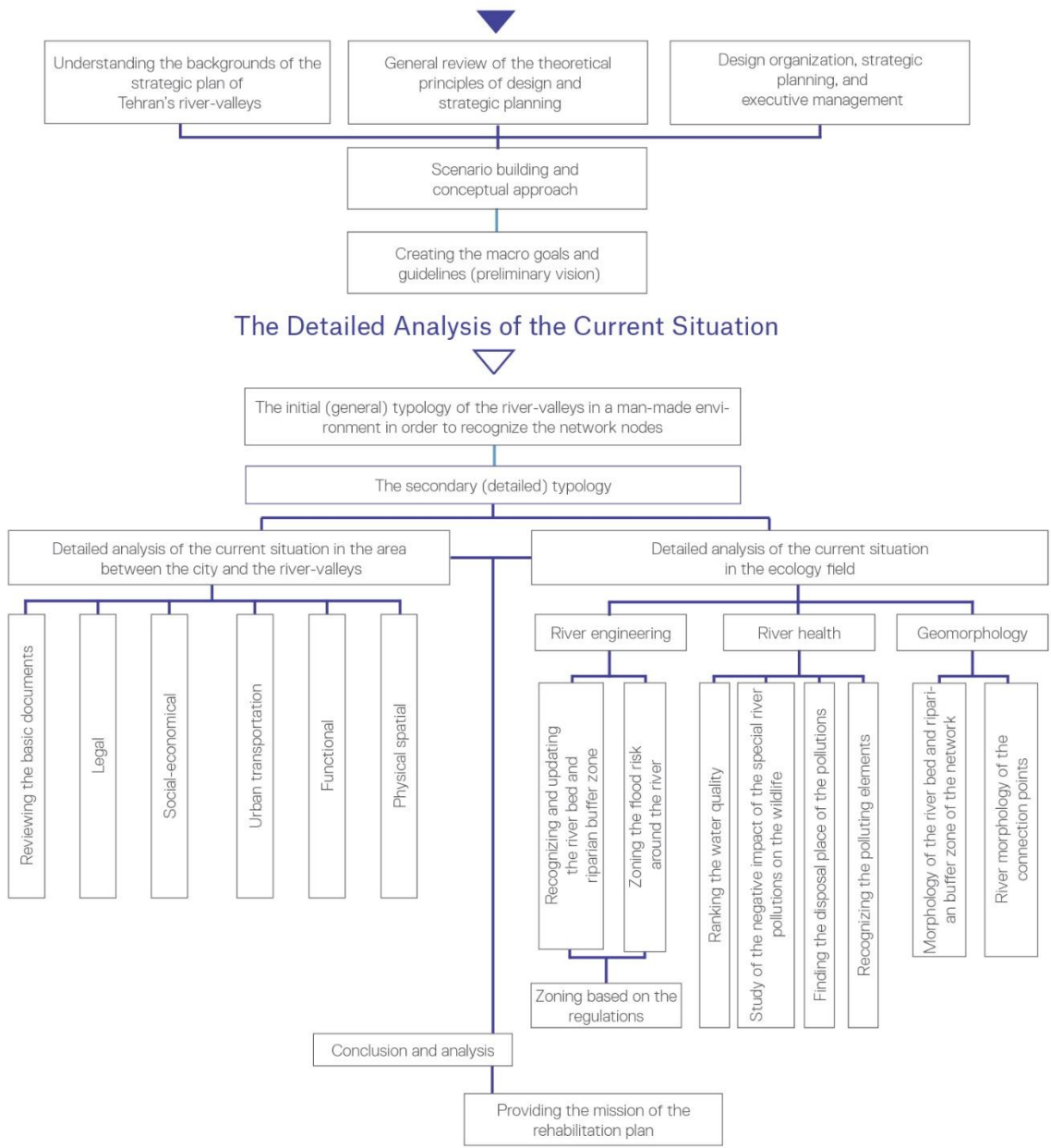

**Diagram 3**: Analysis of the Current Situation

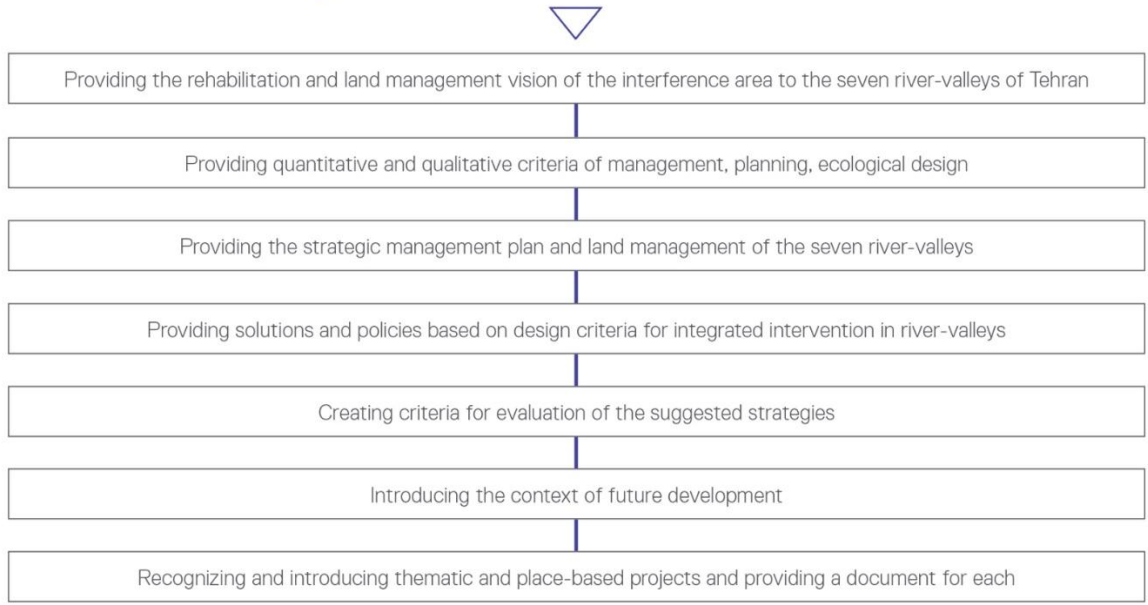

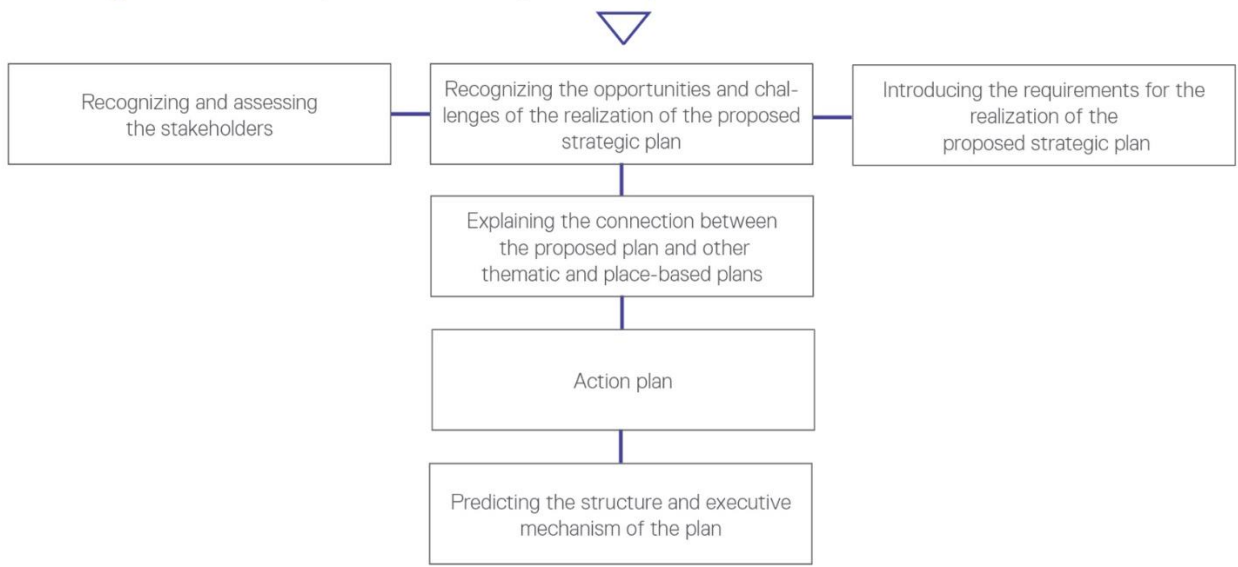

**Diagram.4**: The process

The results of this research have been developed as a guideline to illustrate therehabilitation of the Darband and Darabad river valleys in the following sections:
1. Creating public spaces and providing a space for human interactions and nature
2. Enhancing and maintaining Tehran's natural environment
3. Improving the quality of life of the residents by providing them with an interaction with nature
4. Restoration of floodplains in the long term
5. Enabling the surface runoff in the rivers and improving the surface waters and the distribution of the underground waters
6. The possibility of restoration of the hyporheic zone as a border between the land and the water in the Riverhead, the border between the river and the underground waters, and restoring the river-front vegetation, the wildlife, and improving the river-valley ecotone
7. Creating green river lines and attaching them to the surrounding natural environment
8. Increasing the biodiversity of the river valleys
9. Reducing the noise, visual, and air pollution
10. Creating a new micro-climate using green areas
11. Providing a context for restoring qanats in the long term

**Discussion and conclusion**

This paper has presented an **ecological approach to river rehabilitation** with a focus on the Darband and Darabad valleys in Tehran, and their broader impacts on the urban environment. The growing academic interest in this subject, as reflected in the rising number of publications in recent decades, underscores the critical importance of urban rivers. This trend highlights the emergence of an **interdisciplinary research field** at the intersection of architecture, landscape, and environmental planning. Ecological approaches to urban river rehabilitation have gained particular attention as cities confront challenges such as flooding, sewage mismanagement, visual degradation, and ecosystem disruption. While many of these issues are context-specific, they resonate with urban challenges faced by cities around the world, making the findings of this study broadly relevant.

From a global perspective, the continuous growth and expansion of cities have transformed natural systems at their peripheries into **urban landscape structures**. These "landscape structures of cities" now require careful integration into high-level, urban-scale decision-making frameworks. The blending of "landscape" and "urban" dimensions holds the potential to create synergies between human and natural systems, yet it also brings new layers of complexity and challenges [18].

Recent research highlights the role of **urban ecological networks** in improving environmental quality, conserving natural resources, and fostering human–nature connections. Properly integrated, ecological corridors such as river valleys function as vital linking agents that merge natural and urban systems, creating multifunctional landscapes with ecological, cultural, and social value [19].

Accordingly, another aim of this paper was to explore how urban natural structures, particularly river valleys, can be utilized in ways that foster **mutually supportive human–nature interactions**. Establishing such a complementary relationship enables cities to address both ecological and social needs simultaneously. This approach not only strengthens urban resilience but also advances the pursuit of **sustainability in its fullest sense**.

The case study of Tehran, with emphasis on the Darband river valley, illustrates how urban rivers can serve as vital ecological and cultural assets when reintegrated into the urban fabric. By recognizing their potential as life-giving corridors rather than as drainage systems, cities can transform their rivers into engines of sustainability, biodiversity, and human well-being.

To transform the Darband River Valley into a continuous landscape corridor capable of fostering **win–win human–nature–nature interactions** along its entire 33 km path, this paper proposes a **complementary and continuous scenario** of possibilities. This vision includes the distribution of mountainous recreational facilities in both the northern and southern sections of the valley, ensuring that recreational targets are defined at both the starting point in the north and the endpoint in the south. Such continuity is essential to maintaining the ecological and social integrity of the river valley as a unified system.

Throughout this continuum, the incorporation of the river valley into **urban-scale landscape projects** becomes vital. Accordingly, the proposed scenario emphasizes the integration of the Darband River Valley with the design of existing and potential parks of various scales, ranging from neighborhood to metropolitan levels, across both urban and peri-urban zones of Tehran. These parks provide opportunities for **human–nature as well as human–human interactions** at multiple scales, spanning everyday routines to collective cultural and recreational activities.

Beyond large-scale recreational projects, the Darband River Valley also holds the potential to be embedded within the design of **urban spaces**, such as squares, streets, and bridges located at different points along its course. These elements can serve as critical nodes for reinforcing human–nature connections while enhancing urban identity.

In this complementary scenario, the paper also introduces the concept of **"landscape in transit"**, complemented by strategically designed pause points. Together, these elements create a spectrum of possibilities for interaction with the river valley, forming an integrated system where both stationary and flowing water generate dynamic opportunities for ecological restoration, cultural vitality, and sustainable urban development.[20].

The proposed complementary framework for **human–nature interaction** in the Darband River Valley can be summarized into five main categories:
1. **Inner-city local and medium-sized parks** that foster neighborhood-scale communication and social cohesion.
2. **Urban spaces such as squares and streets**, where the historical identity of the river landscape can be revitalized and reconnected with daily urban life.
3. **Large parks and undeveloped open lands**, which host collective cultural activities and reflect the social attitudes associated with specific days and events.
4. **Mountainous recreational areas**, providing opportunities for leisure, ecological appreciation, and climate-sensitive urban recreation.
5. **Bridges as landscapes in transit**, complemented by designated pause points, create moments of reflection and interaction within the continuum of urban mobility.

Beyond these categories, the Darband River Valley has the potential to engage with the **routines and rituals of Tehran's residents** in numerous other ways. Such interactions depend on the detailed characteristics of each river-adjacent neighborhood and the distinctive practices of its residents. Recognizing this variability opens a new platform for further research aimed at scaling the study from the **urban level to the neighborhood level**.

A comprehensive approach that combines **top-down urban planning strategies** with **bottom-up community engagement** would provide a holistic framework. This integration could yield a full spectrum of design and planning strategies capable of advancing **sustainability in its most complete sense**, ensuring ecological resilience while supporting cultural identity and community well-being [21].

The proposed guidelines and strategies serve as an upstream planning framework for the areas adjacent to Tehran's intermittent streams. A central component of this framework is the creation of **leisure and touristic spaces** along the stream margins, integrating ecological restoration with urban life.

The rehabilitation plan for the Darband and Darabad river valleys, when considered in relation to their surrounding urban fabric, envisions the formation of **urban public spaces** and an **active green corridor** functioning simultaneously as a leisure, cultural, and sports zone. Within this framework, the Darband and Darabad corridors will establish a dynamic local green line that supports the public spaces of Shariati Street and links these areas with cultural zones through a diverse network of parks, pedestrian pathways, urban plazas, and community services.

The introduction of **local pedestrian passages** adjacent to these intermittent streams is expected to reduce vehicular traffic while improving accessibility to residential areas. This integrated pedestrian network will create a more sustainable and people-centered mobility system. Moreover, the incorporation of **clean water as a primary design element** will provide unique and enjoyable urban experiences along the pedestrian routes, strengthening the sensory and ecological qualities of the river valleys[22].

By implementing this strategy, the quality of urban spaces adjacent to the intermittent streams will be significantly enhanced, while the **connectivity between urban nodes** will be reinforced. Collectively, these measures represent a step toward the sustainable integration of rivers into the city's spatial structure, reinforcing their role as life-giving corridors for both nature and society.

In summary, this paper has demonstrated how the **interdependency of different urban sectors** can generate cascading impacts, with losses in one sector contributing to vulnerabilities in others. To address these risks, reliable data on spatial characteristics and land-use properties is essential for estimating **flood-related economic losses** and for providing policymakers with accurate insights to guide risk management strategies.

Scientific evaluations, including **hydrological modeling and cost–benefit analysis**, have proven to be highly effective tools in protecting both property and the environment. The implementation of a **multilateral strategy** enabled decision-makers to identify optimal flood mitigation measures, balancing cost-effectiveness with ecological preservation. Moreover, this strategy highlighted the potential for transforming the **Darband and Darabad river valleys into multifunctional landscapes**, integrating ecological, social, and cultural functions.

The proposed interventions—ranging from targeted improvements and reconstruction to ecological rehabilitation—capitalize on the **existing assets** of the valleys, such as native vegetation, communal spaces, and the engagement of local communities. These measures not only enhance environmental quality but also foster greater public participation in the restoration process.

Ultimately, such initiatives will contribute to:
- Improving the overall **quality of urban life**,
- Attracting tourism and cultural engagement,
- Increasing residents' satisfaction and well-being, and

- Preserving rivers and water systems as fundamental elements of urban sustainability.

**Disclosure statement**

Funding: This research received no external funding.
Clinical trial number: Not applicable.
Consent to Publish declaration: Not applicable.
Ethics and Consent to Participate declarations: Not applicable.
Author Contributions: Both authors contributed to the conception, design, and writing of the manuscript. Specifically, P.B. led the systematic literature review and data analysis, while SH.M. contributed to the drafting, editing, and critical revision of the manuscript. Both authors approved the final version for submission.
Competing Interests: The authors declare no competing interests.

# References

[1] E. Zebardast, "The Application of Analytic Network Process (ANP) in Urban and Regional Planning," *Honar-Ha-Ye-Ziba Memari-Va-Shahrsazi,* vol. 2, no. 41, pp. 79-90, 2010.

[2] E. Zebardast, "Application of F'ANP in Urban Planning," *Honar-Ha-Ye-Ziba Memari-Va-Shahrsazi,* vol. 19, no. 2, pp. 23-38, 2014.

[3] V. M. Masihi, Strategic planning and its application in Iran's urban planning and the case study of Tehran metropolis, Tehran: Urban planning and processing company, 2005.

[4] I. W. R. Management, "Study Services' Details of Identifying Bed Limits of Rivers and Intermittent," *Energy Ministry Journal,* vol. 251, 2003.

[5] S. Manshour and S. Lehmann, "A systematic review of passive cooling strategies integrating traditional wisdom and modern innovations for sustainable development in arid urban environments," *arXiv preprint* arXiv:2507.09365, Jul. 12, 2025. doi: https://doi.org/10.48550/arXiv.2507.09365

[6] H. Ghayor, M. r. Kaviani and B. Mohseni, "Estimation of the coverage level and the amount of snow fall in the northern heights of Tehran, a case study: Djirish river basin (Darband and Golabdera).," *Geographical research,* vol. 19, no. 4, pp. 15-33, 2004.

[7] G. B. Gholikandi, S. Haddadi, E. Dehghanifard and H. R. Tashayouie, "Assessment of surface water resources quality in Tehran province, Iran," *Desalination and Water Treatment,* vol. 37, no. 3, pp. 8-20, 2012.

[8] T. Municipality, Zoning Studies of Flood in the Comprehensive Plan of Tehran:, Tehran: Jihad Water and Basins Research Company, 2005.

[9] R. Ghalambor Dezfoli, "Decision support system in urban management of Tehran with emphasis on urban planning," *Danesh Shahr,* vol. 311, pp. 35-71, 2015.

[10] J. Mehdi Zadeh and M. H. Pirzadeh Nahoochi, Strategic Planning of Urban Development, Tehran: Arman Shahr, 2003.

[11] M. Niyyati, Z. Lasjerdi, M. Nazar, A. Haghighi and E. Nazemalhosseini Mojarad, "Screening of recreational areas of rivers for potentially pathogenic free-living amoebae in the suburbs of Tehran, Iran," *Water & Health,* vol. 10, no. 1, pp. 140-146, 2012.

[12] M. Moosakhaani, L. Salimi, M. T. Sadatipour and M. Rabbani, "Developing Flood Economic Loss Evaluation Model in Residential and Commercial Sectors Case Study: Darband and Golab Darreh Rivers," *Environmental Energy and Economic Research (EEER),* vol. 4, no. 3, pp. 215-229, 2020.

[13] . M. J. Ismaeal and Y. Nazanin, "The Analysis Of Water Crisis Conjecture In Iran And The Exigent Measures For Its Management," *TREND (TREND OF ECONOMIC RESEARCH) ,* vol. 21, pp. 117-144, 2014.

[14] Z. Ebrahimi and S. M. Mahmoudzadeh, "Strategic Orientation of Tourism Sustainable Development: Case Study of Darband," *ournal of History Culture and Art Research focuses on History, Culture and Art,* vol. 6, no. 1, pp. 209-217, 2017.

[15] S. Shobeiri, "Complementary Possibilities in Human-Nature Interaction in an Urban Context – a case study: Darband River-Valley in Tehran," *European Journal of Engineering Science and Technology,* vol. 2, no. 3, pp. 31-52, 2019.

[16] M. S. Foomani and B. Malekmohammadi, "Site selection of sustainable urban drainage systems using fuzzy logic and multi-criteria decision-making," *Water and environmental journal,* vol. 34, no. 4, pp. 584-599, 2019.

[17] M. Tavassoli and N. Bonyadi , Urban space design: urban spaces and their place in city life and appearance, Tehran: Iran Urban Planning and Architecture Studies and Research Center, 2004.

[18] A. Galletta, Mastering the Semi-Structured Interview and Beyond from Research Design, New York : New York University Press., 2013.

[19] S. Manshour, "Utilizing the potential of public spaces for the development of urban leisure and tourism areas (Case study: Zargandeh Neighborhood Central District)," *Current Opinion*, vol. 5, no. 1, pp. 1073–1095, 2025.

[20] B. Abadi, "Could farmers' awareness of environmental NGOs be associated with water conservation behavior? An application of Contingency Table Analysis," *Azarian Journal of Agriculture (AJA),* vol. 4, no. 4, pp. 95-109, 2017.

[21] S. Khatibi and H. Arjjumend, "Water Crisis in Making in Iran," *Grassroots Journal of Natural Resources,* vol. 2, no. 3, pp. 45-54, 2019.

[22] M. Ehsani and H. Khaledi, "griculture water productivity in order to supply water andfood security.," in *1th Seminar of Iranian National Committee on Irrigation and Drainage*, Tehran, 2003.